# Phonon-related monochromatic THz radiation and its magneto-modulation in 2D ferromagnetic Cr$_2$Ge$_2$Te$_6$

*Long Cheng[1,2], Huiping Li[3], Gaoting Lin[4], Jian Yan[4], Lei Zhang[1], Cheng Yang[4], Wei Tong[1], Zhuang Ren[1,2], Wang Zhu[1], Xin Cong[5], Jingjing Gao[4], Pingheng Tan[5], Xuan Luo[4*], Yuping sun[1,4,6], Wenguang Zhu[3*], Zhigao Sheng[1,6*]*

[1] Anhui Key Laboratory of Condensed Matter Physics at Extreme Conditions, High Magnetic Field Laboratory, HFIPS, Anhui, Chinese Academy of Sciences, Hefei 230031, China

[2] University of Science and Technology of China, Hefei 230031, China.

[3] ICQD, Hefei National Laboratory for Physical Sciences at the Microscale, and Key Laboratory of Strongly-Coupled Quantum Matter Physics, Chinese Academy of Sciences, School of Physical Sciences, University of Science and Technology of China, Hefei 230026, China

[4] Key Laboratory of Materials Physics, Institute of Solid State Physics, HFIPS, Chinese Academy of Sciences, Hefei, 230031, China

[5] State Key Laboratory of Superlattices and Microstructures, Institute of Semiconductors, Chinese Academy of Sciences, Beijing 100083, China

[6] Collaborative Innovation Center of Advanced Microstructures, Nanjing University, Nanjing, China

Searching multiple types of terahertz (THz) irradiation source is crucial for the THz technology. In addition to the conventional fermionic cases, bosonic quasi-/particles also promise energy-efficient THz wave emission. Here, by utilizing a two-dimensional (2D) ferromagnetic $Cr_2Ge_2Te_6$ crystal, we firstly demonstrate a phonon-related magneto-tunable monochromatic THz irradiation source. With a low-photonic-energy broadband THz pump, a strong THz irradiation with frequency ~0.9 THz and bandwidth ~0.25 THz can be generated and its conversion efficiency could even reach 2.1% at 160 K. Moreover, it is intriguing to find that such monochromatic THz irradiation can be efficiently modulated by the magnetic field below 160 K. According to both experimental and theoretical analyses, the emergent THz irradiation is identified as the emission from the phonon-polariton and its temperature and magnetic field dependent behaviors confirmed the large spin-lattice coupling in this 2D ferromagnetic crystal. These observations provide a new route for the creation of tunable monochromatic THz source which may have great practical interests in future applications in photonic and spintronic devices.

## Introduction

Terahertz (THz) radiation source plays a crucial role in relevant research and applications. After decades of intensive cultivation, the playground of THz sources has been widely developed in many materials, such as semiconductors,[1] nonlinear electro-optic crystals,[2] surface plasma,[3,4] metamaterials,[5] ferro-/non-magnetic heterojunctions,[6] *etc*. Recently, with extraordinary electrical and optical properties, 2D van der Waals (vdWs) materials has been also used for the novel THz irradiation source developing.[7] For instance, Xu *et.al.* reported the THz surface emission based on the competition between surface optical rectification and photocurrent surge in layered $MoS_2$.[8] Under the excitation of strong infrared pulses, Ma *et.al.* proposed a new THz emitter based on the shift current occurring on the inversion-broken surface of layered $CrSiTe_3$ crystal.[9] With the aid of laser excited surface plasmon on gold substrate, Bahk *et.al.* observed an enhanced THz emission from the single-layered graphene.[10] In the majority of these studies, THz radiation based on 2D vdWs materials are originating from photoconductive effect, optical rectification, laser induced plasma, which all belong to the fermion (electron) catalog and generally bring broadband THz emission with low conversion efficiency.

In addition to the broadband cases, monochromatic THz sources is also important and it possesses immense practical applications potential,[11,12] especially in communication technology.[13] According to the protocol of the International Telecommunication Union (ITU), the better monochromaticity of the waves in communication usually means higher band utilization and lower interference risk.[14] From visible light to near-infrared band, the monochromatic radiation sources (MRS) are usually realized by exciting the electronic transition between the direct band gap of specific semiconductors.[15-18] While to the extent of MRS in mid- and far- infrared range, the radiation sources would extend from fermionic particles to bosonic cases, such as phonons.[19-23] For THz wave with lower photon energy, its MRS requires phonons with matched frequency, which usually could be observed in some artificial materials. For example, by designing proper semiconductor superlattice structures, people have observed the THz phonons and experimentally realized quasi-monochromatic THz radiations.[24,25]

2D vdWs materials are natural superlattice structures with weak *inter*-layered interactions that hundreds of times weaker than *intra*-layered cases.[26] Moreover, layered 2D vdWs materials possess not only high anisotropy but also rich variety of *inter*-layered breathing phonon modes in THz range.[27,28] This feature holds a great potential for the monochromatic THz radiations from the phonons in 2D vdWs materials. Its main advantage, in comparison to fermionic (electron) cases, the phonon related radiations in principle would be 'zero-threshold' and high efficiency with high quality factors due to the its bosonic quasi-/particles nature that is independent of their population and possess relatively long lifetimes.[29-33] Surprisingly, little attention has been devoted to the study of 2D vdWs bosonic THz wave and no cogenetic phonon-based monochromatic THz radiation has been published yet.

Here, by choosing ferromagnet $Cr_2Ge_2Te_6$ as a model system, we firstly demonstrate a 2D phonon-based monochromatic THz radiation. It was found that a strong THz irradiation with frequency ~0.9 THz and bandwidth ~0.25 THz can be generated with a low-photonic-energy broadband THz pump. The conversion efficiency varies with temperature and can even reach 2.1% at 160 K. Moreover, it is intriguing to find that the external magnetic field ($\boldsymbol{B}$) could efficiently modulate such monochromatic THz irradiation especially when $\boldsymbol{B}$ is parallel to the *ab*-plane of the 2D $Cr_2Ge_2Te_6$ crystal. After carefully experimental and theoretical examining, the emergent THz irradiation is identified as the emission from the phonon-polariton in this 2D ferromagnetic crystal. Our findings suggest that the use of 2D materials may provide a variable source for bosonic-related monochromatic THz radiation.

## Results

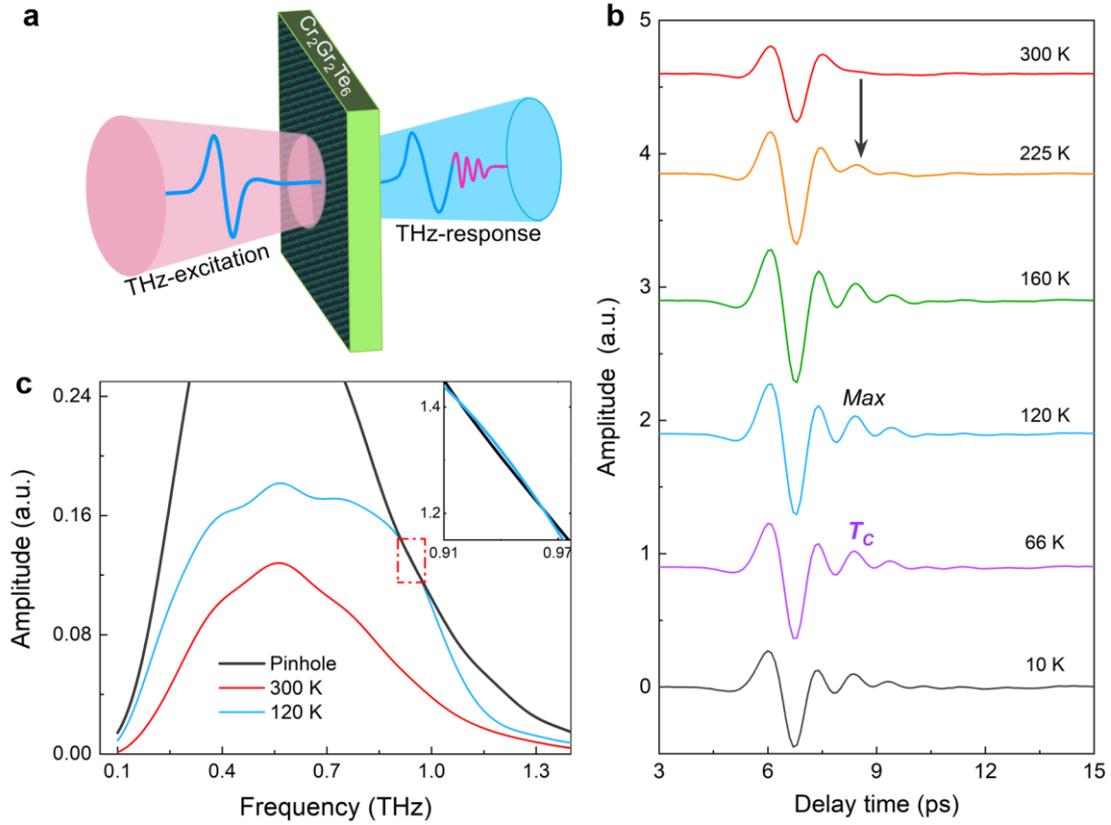

**Fig. 1 | Terahertz responses of Cr₂Ge₂Te₆ crystal. a**, Schematic diagram of the transmission measurement configuration. **b**, Time-resolved THz spectra transmitted thorough the sample at some critical temperatures. **c**, Frequency-domain THz spectra corresponds to temperatures of 300 K (red curve) and 120 K (blue curve). The black curve is the reference THz signal through the pinhole of the sample holder without any sample. The inset is the close-up view of the red dashed frame area at 120 K.

To create a phonon-based THz emitter, appropriate materials are important. In addition to intensively studied metamaterial and graphene, 2D vdWs magnetic materials are another ideal material for this purpose because their exotic properties and their strong spin-phonon coupling are expected to give effective responses to external magnetic fields.[34] $Cr_2Ge_2Te_6$ is one of the prototypes. It undergoes a paramagnetic-ferromagnetic transition at $T_C = 68$ K.[35] The THz transmission response of $Cr_2Ge_2Te_6$ single crystal was measured by means of THz-TDS system as schematically shown in Fig. 1a. The measurements were taken from room temperature to 10 K with an Oxford Spectromag He-bath cryostat (see details in Method part). After penetrating through the sample, the time-resolved THz signals were collected and the typical results are depicted in Fig. 1b. In the time domains of transmitted THz wave obtained at 300 K, there is a main wave which has the same shape as the incident THz wave with lower intensity due to the absorption of $Cr_2Ge_2Te_6$ crystal. With temperature decreasing, the intensity of the main wave increases at first and then drops below 160 K (Supplementary Information Fig. S1). Apart from the main wave, it is intriguing to find that there are additional oscillations superimposed on the rear portion of the main waveforms (black arrow in Fig. 1b). Such THz oscillations emerges at 225 K and exists until the lowest temperature of 10 K we measured. Its amplitude slightly varies with temperature and reaches the maximum at 120 K.

There are two possible origins for these additional oscillations. One is Fabry–Pérot effect that arises from multiple reflections between internal interfaces of the sample, another is electromagnetic (EM) radiations. The loss functions for both experimental results and theoretical predictions of Fabry–Pérot effect were calculated.[36] As shown in the supplementary Fig. S2, the experimental result is significantly different from the theoretical feature of the Fabry–Pérot effect. Especially, the loss function of these oscillations corresponds to the THz response around 0.9 THz and could even be negative, which indicates negative loss after penetrating the sample. In order to further explore the additional oscillations, the fast-Fourier-transformation from TDS results was done and the corresponding typical frequency domain spectra (FDS) results are depicted in Fig. 1c. By comparing the results of 300 K (red curve) and 120 K (blue curve), it is found that the additional oscillations yield a certain frequency around 0.9 THz. Especially, at 120 K, the amplitude of output THz wave at ~0.9 THz is almost 2.7 times of 300 K case and even higher than the free-space pinhole reference (black curve), as shown in the inset of Fig. 1c. Considering both the negative loss function and large output THz wave at ~0.9 THz, it is reasonable to conclude that the additional oscillations are emergent THz radiation.

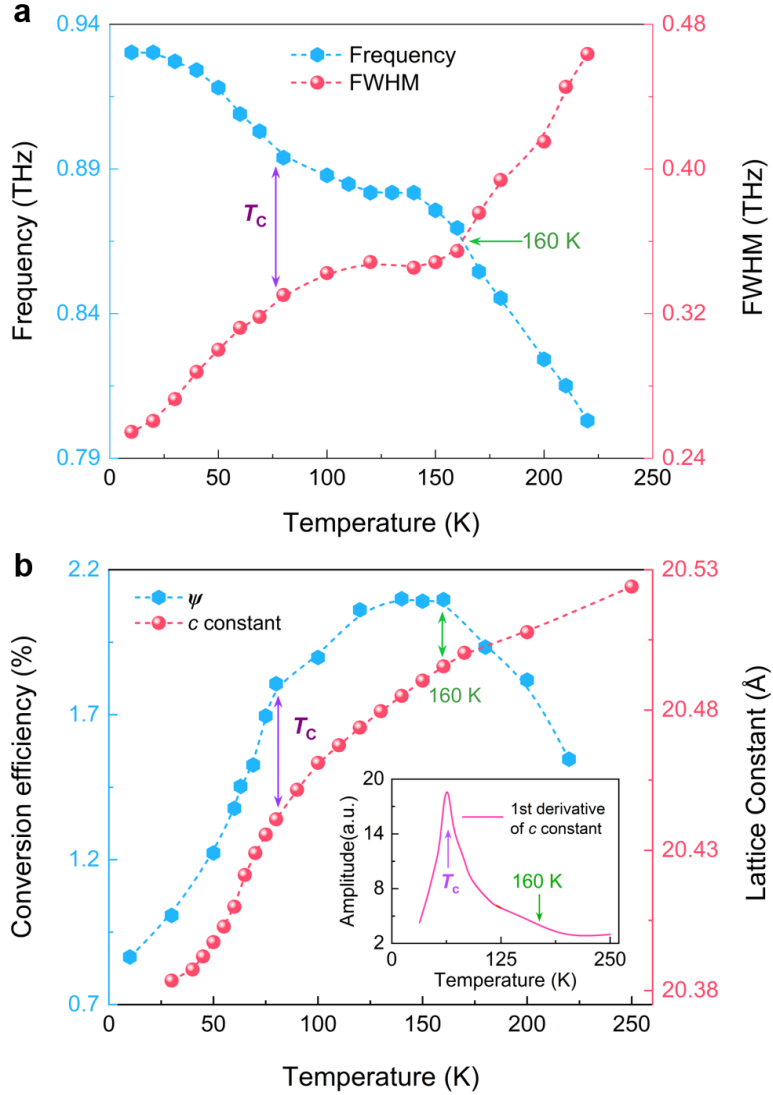

**Fig. 2 | Temperature dependent properties of the radiation and *inter*-layered lattice constant. a**, Temperature dependent center frequency ($f_c$, blue dashed line connected with hexagonal symbols) and FWHM (red dashed line marked with sphere symbols) of the radiation **b**, Temperature dependence of optical-to-THz conversion efficiency ($\psi$) (blue dashed curve marked with hexagonal symbols) and lattice constant along *c*-axis (sphere symbols connected with red dashed curve). These purple and green arrows in **a** and **b** mark the transition temperatures around $T_C$ and 160 K. The inset is the first derivative of the temperature dependent *c* constant.

Furthermore, by calculating the FDS of transmittance variation (*Delta T*) of specific temperature with reference to the case of 300 K that without radiation emerged (Supplementary Information Fig. S3), it is found that the additional oscillation is monochromatic and its center frequency, monochromaticity, and amplitude are strongly dependent on temperature. Figure 2a presents the temperature dependence of the center frequency ($f_c$) and full-width at half-maximum (FWHM) of the THz radiation. At first glance, the most important common performance is that the temperature dependence of $f_c$ and FWHM separate the temperature coordinate into three regions with boundaries of 160 K and $T_C$. While the striking contrast is that their dependences are totally opposite. Specifically, for the case of $f_c$, when temperature drops from 220 K to 160 K, it first dramatically increases from 0.80 to 0.87 THz. While in the following region from 160 K to $T_C$, it turns to increase in a slowing trend from 0.87 to 0.90 THz. Afterwards, at temperatures lower than $T_C$, the frequency is further hardened to 0.93 THz. However, for the FWHM that indicates the monochromatism of the radiations, its temperature dependence in these three regions corresponds to the narrowing from 0.46 to 0.35 THz rapidly (220 K to 160 K), the approximate plateau form from 0.35 to 0.32 THz (160 K to $T_C$), and a further narrowing to 0.25 THz below $T_C$, respectively.

As a rule of thumb, the conversion efficiency ($\psi$) is a critical characteristic.[37] Here, the radiation performance is evaluated with the parameter of $\psi = I_r/I_{ab} = E_O^2 / (E_i^2 - E_t^2)$, where $I_{ab}$ and $I_r$ are intensity of absorbed pump pulse and generated THz radiation, $E_O$ is the P-P amplitude of the first oscillation waveforms, $E_i$ and $E_t$ are P-$P_M$ of the incident and transmitted THz waveforms, respectively. Corresponding results are shown as hexagonal symbols in Fig. 2b. The $\psi$ also shows the same temperature regions as $f_c$ and FWHM. From 220 K to 160 K, the $\psi$ starts to emerge and first be dramatically enhanced to ~2.1%, which is higher than the cases of fermionic cases, such as photoconductive and nonlinear optical rectification (usually lower than 0.1%).[38-40] Whereas in the region below 160 K, the increasing trend then gradually transforms into a declining one. Peculiarly, for temperatures below $T_C$, it is dramatically attenuated and finally reaches ~0.86% at 10 K. All in all, the temperature dependence of $\psi$ possesses the same features and originations as the case of P-$P_M$, in which transformations below 160 K and $T_C$ could attribute to the enhanced THz absorption arising from the formation of magnetic correlations.[41-43] However, the emergency temperature (225 K) of the THz radiation is far away from the Curie temperature ($T_C$ ~68 K) and also above the spin fluctuation temperature (~160 K).[43,44] Moreover, the THz radiation frequency (~0.9 THz/3.7 meV) of $Cr_2Ge_2Te_6$ crystal is quite larger than its spin excitation gap (~0.28 meV).[45] These two factors imply that the observed THz radiation is not entirely spin related.

Previous studies on semiconductor-based THz sources indicate that the THz radiations are usually assisted by intrinsic phonons in these materials.[46-48] As shown above that the frequency, FWHM, and the intensity of the additional THz emission in this 2D material are temperature dependent and there are two critical temperatures (160 K and $T_C$) exist. The X-ray diffraction (XRD) measurements with temperature variation were done and the temperature dependent lattice constant along $c$-axis are shown in Fig. 2b (red spheres). It is found that the $c$-axis between two CGT layers decreases with decreasing of the temperature and the decreasing slopes change twice when sample was cooling from room temperature down to 25 K. As shown by its first order derivative curve in the inset of Fig. 2b, the critical temperatures corresponding to the decreasing slope variations are 160 K and $T_C$, which is same as those for the parameters of the THz emission. Such coincidence implies that the observed THz emission is associated with the *inter*-layered phonons in this 2D vdWs ferromagnetic material.

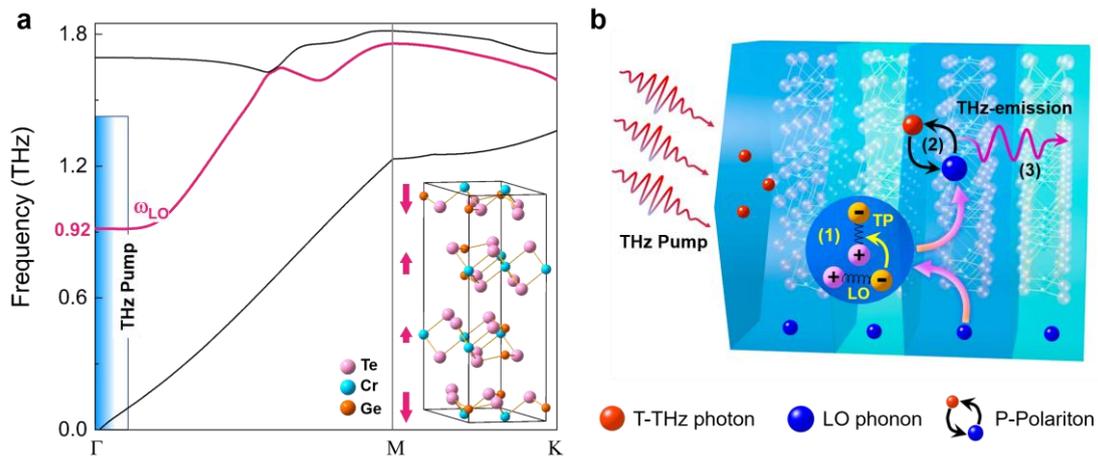

**Fig. 3 | Schematic of phonon-polariton generating THz radiation scheme. a**, Sketched phonon dispersion of $Cr_2Ge_2Te_6$ crystal corresponding to the THz radiation. The THz radiation involved zone-folded LO layer-breathing mode is depicted with red curve. While black curves are radiation irrelevant phonons. The shaded region exhibits the spectral range of incident THz pump. Inset shows the crystal structure of the $Cr_2Ge_2Te_6$ crystal, in which these red arrows indicate the oscillation directions of the *inter*-layered breathing mode. **b**, Diagram of the coupling of LO layer-breathing mode and THz photons, resulting in the formation of P-

Polaritons and ensuing monochromatic THz EM emission. The inset marked with pink arrow illustrates the intermediate (TP) during the whole process. The formation processes of TP, P-Polariton, and THz emission correspond to (1), (2), and (3), respectively

Owing to the unit cell of the $Cr_2Ge_2Te_6$ expanded along $c$-axis, the Brillouin zone is folded in the momentum space, resulting in the emergence of zone-folded phonon modes. Based on the XRD results, we calculated the phonon dispersion of the expanded hexagonal cell with the *phonopy* code interfaced with VASP (see detailed results in supplementary Fig. S4).[49,50] As notably shown with red arrows in the inset of Fig. 3a, there is an *inter*-layered longitudinal vibration mode, corresponding to a non-centrosymmetric longitudinal optical (LO) layer-breathing mode with frequency ($\omega_{LO}$) of ~0.92 THz at $\Gamma$ point. Such LO mode with the three layers within each unit cell possesses different vibration amplitudes (red curve in Fig. 3a). It should be emphasized that this LO layer-breathing mode, which breaks the inversion center and degenerates the space group from $R\bar{3}$ to $R3$, is dipole active and gives rise to piezoelectricity.[51] Interestingly, due to the piezoelectricity and generalized Hooke's law of the updated group, some co-frequency transverse vibrations with transversal polarization (TP) can be driven by the vibration of LO phonons, as schematically shown with the (1) of Fig. 3b. According to the Lyddane-Sachs-Teller (LST) relation[26], after pumping by a THz pulse envelope with transversal polarized electric field, the incident EM waves with frequency close to the LO layer-breathing mode could effectively propagate in $Cr_2Ge_2Te_6$ and strongly coupled with the co-frequency TP and form a new bosonic quasi-particle, i.e., so called phonon-polariton (P-Polariton).[52,53] After that, these photon pumping induced bosonic P-Polariton could in turn generate far-field EM radiation with frequency close to the original LO phonon as schematically illustrated by process (2) and (3) in Fig. 3b.[21,54] Due to the energy of this bosonic P-Polariton generated THz radiation is carried in the piezoelectricity vibrations, it is reasonable to consider it as a highly efficient bosonic laser-like process.[29,55] Beyond that, according to the further calculations (Supplementary Information Fig. S5), it was found that the $c$ constant dependent evolution of phonon frequency is consistent with both the experimental results of $f_c$ shown in Fig. 2a and the previous reports of spin-phonon interactions.[34]

According to the above discussion and the tensor symmetry of $R3$ structure, it can be predicted that, for a certain THz-Pump field, the electric displacement and corresponding THz radiation should possess a 6-fold in-plane symmetry (see Part VI of the Supplementary Information). In order to confirm this issue, the in-plane anisotropic dependence of the THz radiation amplitude was investigated by tuning the in-plane azimuth of the sample. As shown in Fig. S6b, it exhibits a cycle of $\pi/3$ that matches well with the 6-fold symmetry of the P-Polariton as predicted. In addition, the THz excitation amplitude dependence of the radiation amplitude was also investigated by inserting two THz polarizers after the low-temperature-grown GaAs photoconductive antenna (LT-GaAs-PCA). As shown in Fig. S6d, the radiation and excitation amplitudes are linearly correlated and there is no saturation and threshold observed, which is also in good consistent with the feature of bosonic lasing reported before.[30] These features not only indicate that the radiation effect could be further improved by enhancing the stimulation intensity, but also means that the stimulated irradiation source in $Cr_2Ge_2Te_6$ is extremely sensitive to the weak THz field (~kV/cm). This is in good accordance with the extremely high sensitivity of $Cr_2Ge_2Te_6$ to external low-intensity photonic field.[56] Unlike other THz sources based on 2D materials,[8-10] this unique THz-MRS takes full advantage of the *inter*-layered breathing phonons and achieves highly efficient monochromatic THz radiations with stimulation of low-energy THz pulse. It is of great practical interests, such as monochromatic THz laser, integrated gain-type monochromatic THz filter, key relay components in photonic communication, *etc*.

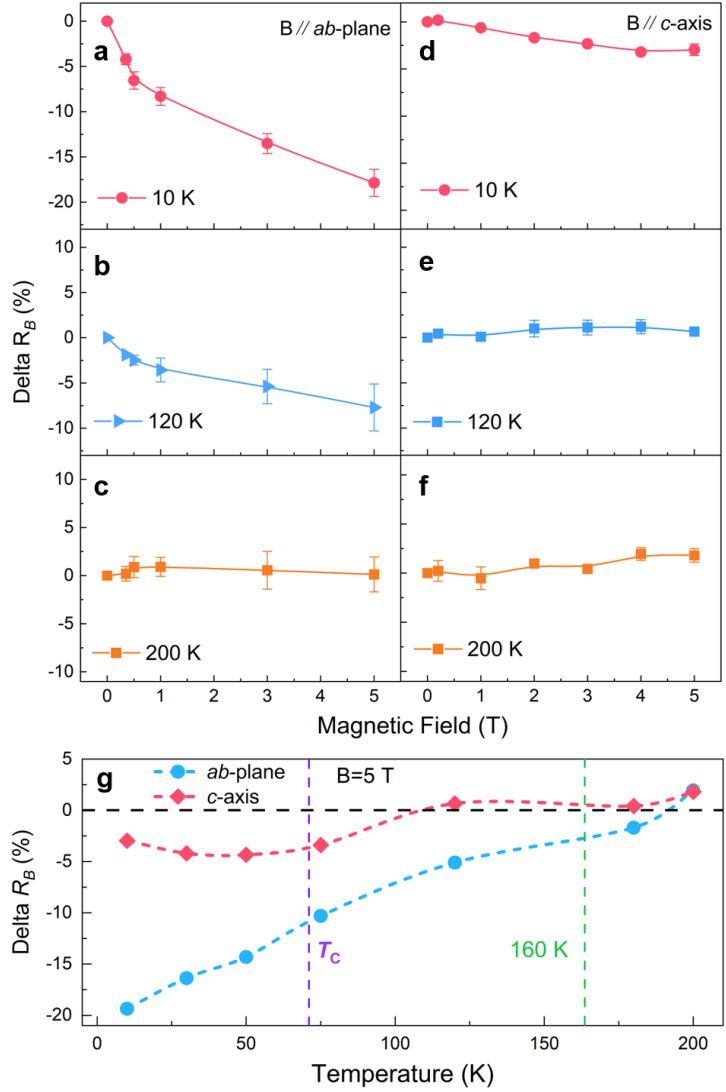

**Fig. 4 | Magnetic field and temperature dependent modulation effect of THz radiation. a-f**, At 10 K, 120 K, and 200 K, $\boldsymbol{B}$ dependence of *Delta* $R_B$ at corresponding $f_c$ with $\boldsymbol{B}$ (from 0 T to 5 T) applied along *c*-axis and *ab*-plane, respectively. Error bars in **a-f** represent uncertainties in determining the *Delta* $R_B$. **g**, Temperature dependent *Delta* $R_B$ under a uniform $\boldsymbol{B}$ of 5 T. Blue and red curves marked with solid circle and square symbols are *Delta* $R_B$ with $\boldsymbol{B}$ applied along *ab*-plane and *c*-axis, respectively. The purple and green dashed lines are boundaries of the temperature regions.

Different from other 2D semiconductors, $Cr_2Ge_2Te_6$ holds ferromagnetism. Except electron and phonon, spin is non-negligible degree and it has been demonstrated that there exist strong spin-phonon coupling in $Cr_2Ge_2Te_6$ and its isostructural $Cr_2Si_2Te_6$ materials.[34,57,58] As presented in Fig. 2, it is found that not only the *c*-axis lattice constant but also the amplitude, frequency, and FWHM of the emergent THz emission have abnormal near around the $T_C$ of $Cr_2Ge_2Te_6$ crystal, which also imply the subtle interplays between the phonon-involved THz radiation and spin orders. Hence, it is reasonable to expected that the THz radiation could also be tuned by external magnetic field ($\boldsymbol{B}$).

The effect of $\boldsymbol{B}$ on the THz radiations was investigated by the THz-TDS system with a superconductor magnet. The system stability under high magnetic field was verified beforehand (see supplementary Fig. S7a). At low temperatures, there is significant variation of the THz response could be observed when $\boldsymbol{B}$ is applied (see supplementary Fig. S7b-d. In order to present the magnetic field effect on the THz emission, the relative THz amplitude variation $Delta\,R_B$ was defined as $Delta\,R_B = (E_B - E_{B0})/E_{B0}$, in which $E_B$ and $E_{B0}$ are THz responses of $Cr_2Ge_2Te_6$ crystal with and without $\boldsymbol{B}$ applied. The magnetic field dependence of $Delta\,R_B$ obtained at corresponding $f_c$ and different temperatures were summarized in the Fig. 4a-f. For $\boldsymbol{B}//c$-axis, the $Delta\,R_B$ is negligible above $T_C$ and it is also small (< -5%) below $T_C$. When $\boldsymbol{B}//ab$-plane, the $Delta\,R_B$ starts to emerge around 160 K and increases with decreasing of the temperature. Notably, at 10 K, the $Delta\,R_B$ in $ab$-plane could reach the maximum around -20% (with $\boldsymbol{B} = 5$ T), which is greater than the case of $c$-axis ($\sim$-3%). Such anisotropic feature of the magnetic field modulation effect on $Delta\,R_B$ can be clearly observed in Fig. 4g, in which the temperature dependent $Delta\,R_B$ under $\boldsymbol{B} = 5$ T is plotted. By applying an external magnetic field, the spin state of the $Cr_2Ge_2Te_6$ crystal can be modified for both long-range order (below $T_C$) and short-range order ($T_C < T < 160$ K).[43,44] As one of the possible mechanism, the magnetic field induced the spin-flips scattering could suppress the electron-phonon interactions as well as the dipole active phonon involved P-Polariton.[59,60] Accordingly, with such large spin-phonon coupling in this 2D ferromagnetic material, the switching of spin state by $\boldsymbol{B}$ would bring the modulation of phonon-related THz emission. Moreover, the spontaneous spin orientation of $Cr_2Ge_2Te_6$ is along $c$-axis. And then, from viewpoint of spin modulation, the magnetic field effect is significant when $\boldsymbol{B}//ab$-plane, which is consistent with our observations shown in Fig. 4. The magneto modulation effect provides another proof for its P-Polariton-related radiation mechanism. All these features not only offer another degree of freedom for the regulation of THz radiation in practical applications, but also provide a convincing evidence for the correlation between the THz radiation, phonons, and spin orders.

## Summary

In summary, benefiting from the coherent LO layer-breathing mode, a magneto-tunable monochromatic THz radiation was demonstrated in a 2D vdWs ferromagnet $Cr_2Ge_2Te_6$. After pumping by a broadband THz wave, a strong monochromatic THz irradiation can be generated. The frequency, FWHM, and the intensity of the emergent THz radiation vary with the temperature and its conversion efficiency could even reach 2.1% at 160 K. Our experimental and theoretical analyses indicate that the emergent THz irradiation is from the pump induced phonon-polariton in this 2D vdWs material. Moreover, due to the existing strong spin-phonon coupling, it is interesting to find that such monochromatic THz irradiation can be efficiently modulated by the magnetic field below 160 K. In addition to the general fermionic cases, our findings suggest that the use of 2D vdWs ferromagnet might provide a viable source for the realization of bosonic THz source with tunable and monochromatic features, which may have great practical interests in future applications in photonic and spintronic devices.

## Methods

The $Cr_2Ge_2Te_6$ single crystals are prepared with the self-flux technique, following the procedure described in ref. [61,62]. The effective size of the sample is ~6×4×0.2 mm. The terahertz data was collected with home-built terahertz time domain spectroscopy system (THz-TDS) has been introduced in detail in ref. [63]. The sample is assembled on the sample holder of Oxford Instruments Spectromag He-bath cryostat, with which one could realize test environments of temperature from 10 to 300 K and magnetic field up to 7 Tesla. After propagating through these fused silica windows of the cryostat, the coverage of the THz spectral is from 0.1 to 1.5 THz. The temperature dependent X-ray diffraction (XRD) was measured with the XRD setup using high-intensity graphite monochromatized Cu Kα radiation system (Rigaku corporation, Japan). Then, the influence of cooling process on the copper base is excluded and finally obtain the accurate temperature dependent evolution of the lattice constant.

## Acknowledgements

We gratefully acknowledge financial support from the National Key R&D Program of China (Grant No. 2017YFA0303603, 2016YFA0401803), the National Natural Science Foundation of China (NSFC; Grant No. U2032218, U1932217, 11574316, 12074386, 11904116, 11874357, U1932217, U1832141), the Plan for Major Provincial Science&Technology Project (Grant No. 202003a05020018), the Key Research Program of Frontier Sciences, CAS (Grant No. QYZDB-SSW-SLH011), China Scholarship Council (CSC). A portion of this work was performed on the Steady High Magnetic Field Facilities, CAS and supported by the High Magnetic Field Laboratory of Anhui Province.

## Author contributions

Long Cheng performed the THz-TDS and other related experiments with help from Zhigao Sheng, Zhuang Ren, Wang Zhu, Jingjing Gao, Xing Cong. Wenguang Zhu and Huiping Li performed the theoretic calculations. Yuping Sun, Xuan Luo, and Gaoting Lin offered high quality samples. The data were analyzed and interpreted by Zhigao Sheng, Long Cheng with help from Wenguang Zhu and Xuan Luo. The manuscript was written by Zhigao Sheng, Long Cheng, Wenguang Zhu and Huiping Li. The project was planned, directed, and supervised by Zhigao Sheng. Long Cheng, Huiping Li contributed equally to this work. All authors discussed the results and commented on the manuscript.

## Additional information

Supplementary information is available in the online version of the paper.

Correspondence and requests for materials should be addressed to Zhigao Sheng.

## Competing financial interests

The authors declare no competing financial interests.

# Supplementary Information for

# Phonon-related monochromatic THz radiation and its magneto-modulation in 2D ferromagnetic semiconductor Cr₂Ge₂Te₆

*Long Cheng[1,2], Huiping Li[3], Gaoting Lin[4], Jian Yan[4], Lei Zhang[1], Cheng Yang[4], Wei Tong[1], Zhuang Ren[1,2], Wang Zhu[1], Xin Cong[5], Jingjing Gao[4], Pingheng Tan[5], Xuan Luo[4*], Yuping sun[1,4,6], Wenguang Zhu[3*], Zhigao Sheng[1,6*]*

[1] Anhui Key Laboratory of Condensed Matter Physics at Extreme Conditions, High Magnetic Field Laboratory, HFIPS, Anhui, Chinese Academy of Sciences, Hefei 230031, China

[2] University of Science and Technology of China, Hefei 230031, China.

[3] ICQD, Hefei National Laboratory for Physical Sciences at the Microscale, and Key Laboratory of Strongly-Coupled Quantum Matter Physics, Chinese Academy of Sciences, School of Physical Sciences, University of Science and Technology of China, Hefei 230026, China

[4] Key Laboratory of Materials Physics, Institute of Solid State Physics, HFIPS, Chinese Academy of Sciences, Hefei, 230031, China

[5] State Key Laboratory of Superlattices and Microstructures, Institute of Semiconductors, Chinese Academy of Sciences, Beijing 100083, China

[6] Collaborative Innovation Center of Advanced Microstructures, Nanjing University, Nanjing, China

# Contents:

# I. Temperature dependence of P-P$_M$

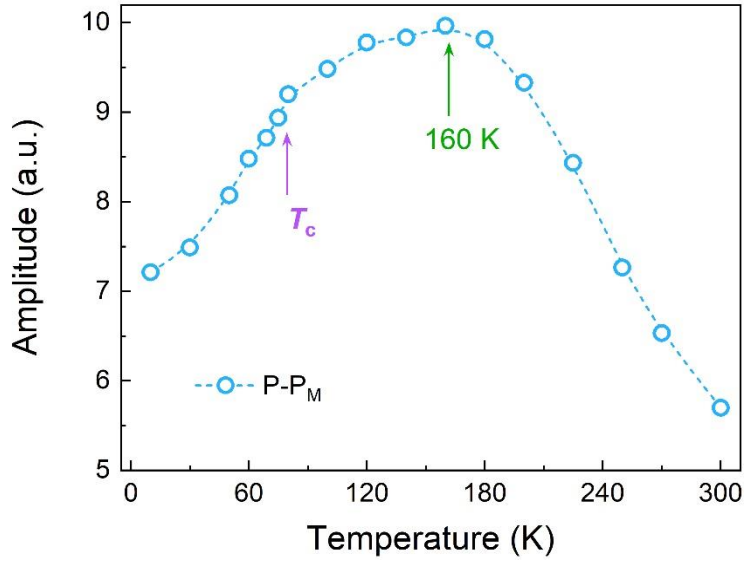

**Fig. S1 | Temperature dependent peak-to-peak amplitude of the transmitted THz main waveforms (P-P$_M$).**

According to the time-resolved spectra of the transmitted THz signals, the peak-to-peak amplitude of the main waveforms (P-P$_M$) could be easily picked out. Its temperature dependence is depicted in Fig. S1. The evolution of the P-P$_M$ divided the temperature coordinate into three regions with the boundaries of 160 K and $T_C$. To be specific, for the main waveforms in time domain, as the temperature drops from 300 K, the P-P$_M$ was significantly enhanced until 160 K. Afterwards, along with the temperature further dropped in the range of 160 to 120 K, the upward trend of the P-P$_M$ slows down and reaches the maximum. Peculiarly, the P-P$_M$ enhancement of 120 K with respect to the 300 K case could reach ~40%. At last, the trend eventually develops into a steep declining one below $T_C$ and reaches the minimum at 10 K. According to previous studies on THz responses of semiconductor materials, the enhancement of transmittance at temperature above 160 K could attribute to the attenuated transition of intrinsic carriers among intra-band levels.[1,2] As to the temperature below 160 K, magnetic correlations start to appear and ensuing spin-electron interactions come into play, which dramatically give rise to an absorption in THz range.[3-5] As a result, during the cooling process, the competition of these two mechanisms would subsequently bring about the transition and decline of the P-P$_M$ amplitude.

## II. Experimental and theoretical loss function

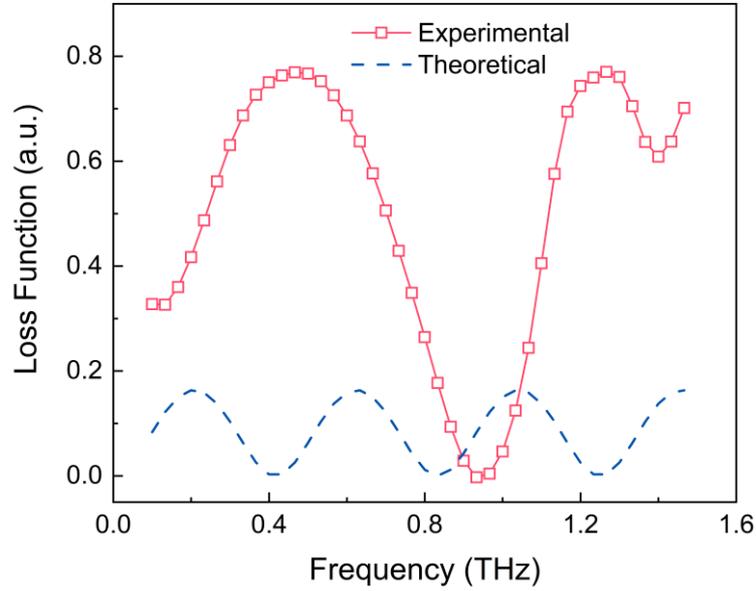

**Fig. S2 | Loss function of experimental (red square) and theoretical (blue dashed) results at 120 K.** All these results are calculated according to ref. [6] The experimental results merely corresponds to the THz response at ~0.9 THz, which is significantly different from the characteristic of Fabry–Pérot (FP) effect that originates from multiple reflections of the THz pulse between internal interface of the sample (shown as dashed curve). Especially, the loss function of these oscillations corresponds to the frequency around 0.9 THz and could even be negative at 120 K, which indicates negative loss after penetrating the sample.

# III. Frequency domain spectra of *Delta T*

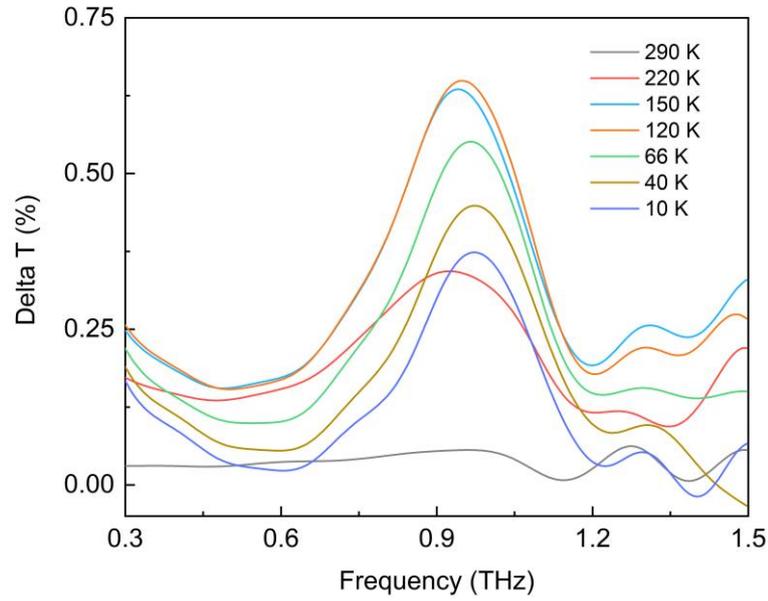

**Fig. S3 | At specific temperature, frequency domain spectra (FDS) of the transmittance variation (*Delta T*) with reference to the case of 300 K.** The results are calculated according to the formula of *Delta T*=$(E_T-E_0)/E_r$, where the $E_T$ and $E_0$ are FDS of the sample at specific and room temperature, respectively. While $E_r$ is the FDS of THz pulse penetrate through the pinhole of the sample holder without $Cr_2Ge_2Te_6$ mounted. This processing method can make the performance of the radiation effect in FDS more intuitive and significant.

## IV. Calculation results of phonon dispersion

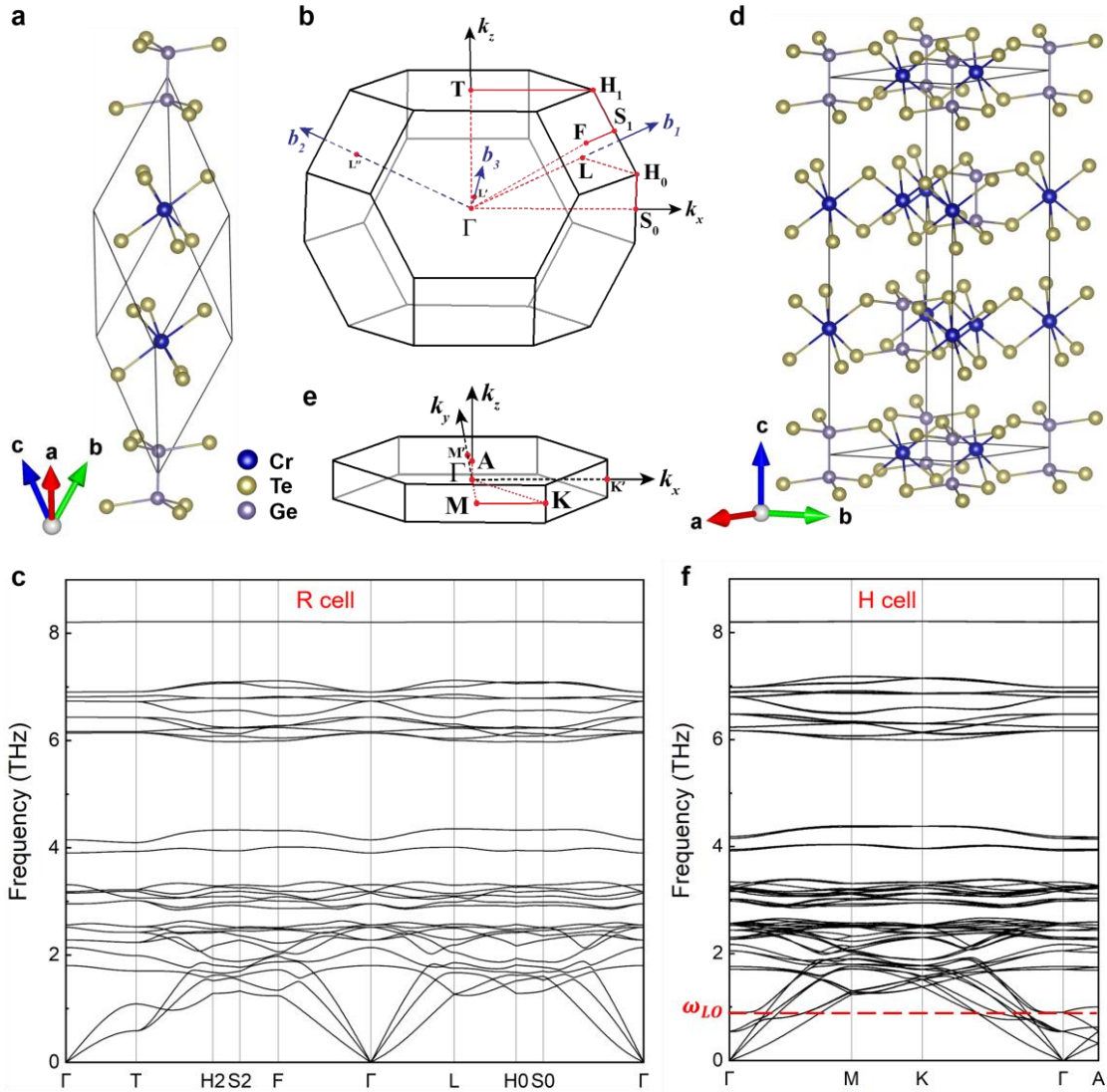

**Fig. S4 | The crystal structure, Brillouin zone, and phonon dispersion of Cr₂Ge₂Te₆ crystal. a**, The rhombohedral primitive cell (R cell) of Cr₂Ge₂Te₆. Blue, purple, and golden spheres represent Cr, Ge, and Te atoms, respectively. **b**, The Brillouin zone corresponds to the R cell. **c**, Phonon dispersion of the R cell. **d**, Crystal structure of Cr₂Ge₂Te₆ in a hexagonal conventional cell (H cell). **e**, Brillouin zone of the H cell. **f**, Zone-folded phonon dispersion of the H cell. The LO layer-breathing mode with frequency ($\omega_{LO}$) around 0.92 THz is marked with red dashed line. All these calculations

10  are carried out with *phonopy* code interfaced with VASP.[7,8]

11      According to the inter-layered lattice constant derived from XRD results, the crystal

12  structure of $Cr_2Ge_2Te_6$ is elongated along *c*-axis.[9] Consequently, the unit cell of

13  $Cr_2Ge_2Te_6$ would transform from rhombohedral primitive cell (R cell) to hexagonal cell

14  (H cell), which induces the folding of the Brillouin zone and emergence of zone-folded

15  phonon mode in the momentum space.[10] Corresponding the phonon dispersion of

16  original R cell and the expanded H cell were calculated with the *phonopy* code

17  interfaced with VASP, [7,8] as shown in Fig. S5.

19  **V.  Estimation of longitudinal optical (LO) phonon**

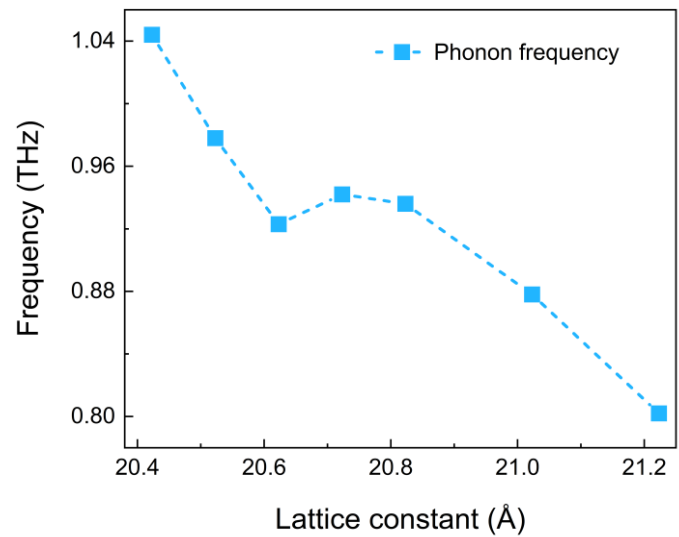

21  **Fig. S5 | Estimation of the spin-phonon interaction induced frequency evolution of**

22  **the inter-layered longitudinal optical (LO) phonon mode with respect to the lattice**

23  **constant.** All these calculations are carried out with *phonopy* code interfaced with

24  VASP.[7,8]

## VI. In-plane anisotropy and stimulation dependence

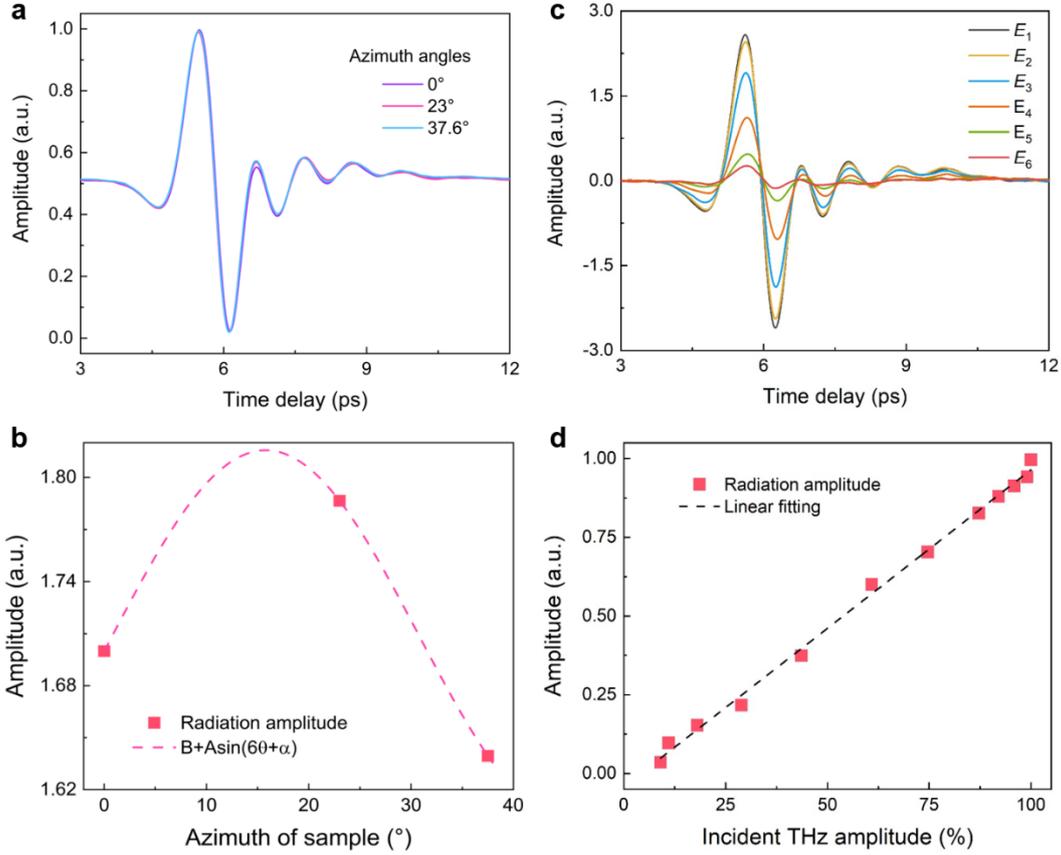

**Fig. S6 | In-plane azimuth and incident THz amplitude dependence of the radiation effect at 120 K. a**, Normalized time-resolved spectra of $Cr_2Ge_2Te_6$ with in-plane azimuth of the sample rotated from zero to 37.6°. **b**, Sample azimuth dependence of the radiation amplitude. These red dashed line in **b** is sinusoidal fitting results for in-plane anisotropy. **c**, Time resolved spectral of $Cr_2Ge_2Te_6$ with incident THz amplitude of the main waveforms decreased from $E_1$ to $E_6$. All these $E_n$ are P-P amplitude of the incident THz pulse. **d**, Incident THz amplitude dependence of radiation amplitude. These percentages of $x$-coordinate are defined with $E_n/E_1$. The dashed line is the linear fitting result.

In this part, the in-plane anisotropy and incident THz amplitude dependence of the radiation effect is investigated by performing the in-plane azimuth dependence. For the case of variating azimuth of the sample, due to the degenerated crystal point group of $R3$, it's sufficient to choose three azimuth angles within the symmetry period ($2\pi/3$) to verify the in-plane anisotropy. As shown in Fig. S6a and b, for these three different in-plane azimuth angles, they also demonstrate certain azimuth dependence. The sinusoidal fitting result yields the 6-fold symmetry that twice of the crystal case. While for the incident THz amplitude dependence of the radiation effect, the radiation amplitude is linearly dependent with respect to the amplitude of the incident THz-pump field.

Given the non-centrosymmetric dipole-active LO phonon mode of the $Cr_2Ge_2Te_6$ crystal, its piezoelectric (PE) performance under external electric field could be expressed as:

$$S_i = s_{ij}\sigma_j + d'_{ij}E_j \tag{S2}$$

$$D_i = d_{ij}\sigma_j + \varepsilon_{ij}E_j \tag{S3}$$

where the $s_{ij}$, $d_{ij}$, $d'_{ij}$, $S_i$, $\sigma_j$, and $\varepsilon_{ij}$ are simplified tensors of compliance, direct PE effect, converse PE effect, phonon involved strain and stress, and permittivity, respectively. While $D_i$ and $E_j$ are vectors of electric displacement and electric field.[11-13] Specifically, for the case of degenerated $R3$ group, $\varepsilon_{ij}$, $d_{ij}$, and $s_{ij}$ for $Cr_2Ge_2Te_6$ could be expressed as:[14]

$$\boldsymbol{\varepsilon_{ij}} = \begin{bmatrix} \varepsilon_{11} & 0 & 0 \\ 0 & \varepsilon_{11} & 0 \\ 0 & 0 & \varepsilon_{33} \end{bmatrix} \tag{S4}$$

$$d_{ij} = \begin{bmatrix} d_{11} & -d_{11} & 0 & d_{14} & d_{15} & -d_{22} \\ -d_{22} & d_{22} & 0 & d_{15} & -d_{14} & -d_{11} \\ d_{31} & d_{31} & d_{33} & 0 & 0 & 0 \end{bmatrix} \tag{S5}$$

$$s_{ij} = \begin{bmatrix} s_{11} & s_{12} & s_{13} & s_{14} & -s_{25} & 0 \\ s_{12} & s_{11} & s_{13} & -s_{14} & s_{25} & 0 \\ s_{13} & s_{13} & s_{33} & 0 & 0 & 0 \\ s_{14} & -s_{14} & 0 & s_{44} & 0 & s_{25} \\ -s_{25} & s_{25} & 0 & 0 & s_{44} & s_{14} \\ 0 & 0 & 0 & s_{25} & s_{14} & \frac{1}{2}(s_{11} - s_{12}) \end{bmatrix} \tag{S6}$$

Then, according to previous studies, the components of the compliance tensor possess the values in the same magnitude of $s_{11} = 14.1 \times 10^{-12} \, \text{Pa}^{-1}$ and $s_{13} = -5.1 \times 10^{-12} \, \text{Pa}^{-1}$, whose difference is much smaller than other 2D materials.[15] As a result, the mechanical vibration of the LO phonon could result in considerable co-frequency transverse strain ($S_1 = s_{13}\sigma_3$) and corresponding piezoelectric displacement ($D_{PE} = d_{11}\sigma_1$). Moreover, by coupling with the co-frequency part of incident transversal envelop of THz-pump field ($E_1$), they could be further coherently enhanced and yield stronger transversal vibrations ( $S_1 = s_{13}\sigma_3 + d'_{11}E_1$ ) and electric displacement ($D_1 = D_{PE} + D_{THz-pump} = d_{11}\sigma_1 + \varepsilon_{11}E_1$), which is also the formation process of a new bosonic particle, namely phonon-polariton (P-Polariton). In addition, from the equation of EqS3, under a certain THz-Pump field, we can find the electric displacement is proportional to the product of $d_{ij}$ and $\sigma_j$. Accordingly, the 6-fold symmetry shown in Fig. S4d could be attributed to the 3-fold of the PE tensor ($d_{ij}$) multiplied by the 2-fold of the phonon involved stress ($\sigma_j$). While for a certain azimuth angle and phonon mode, the linear dependence of the radiation on the incident THz amplitude corresponds to the second term on the right side, which indicates $D_i \propto \varepsilon_{ij}E_j$.

## VII. System stability verification and *B* modulation of THz response

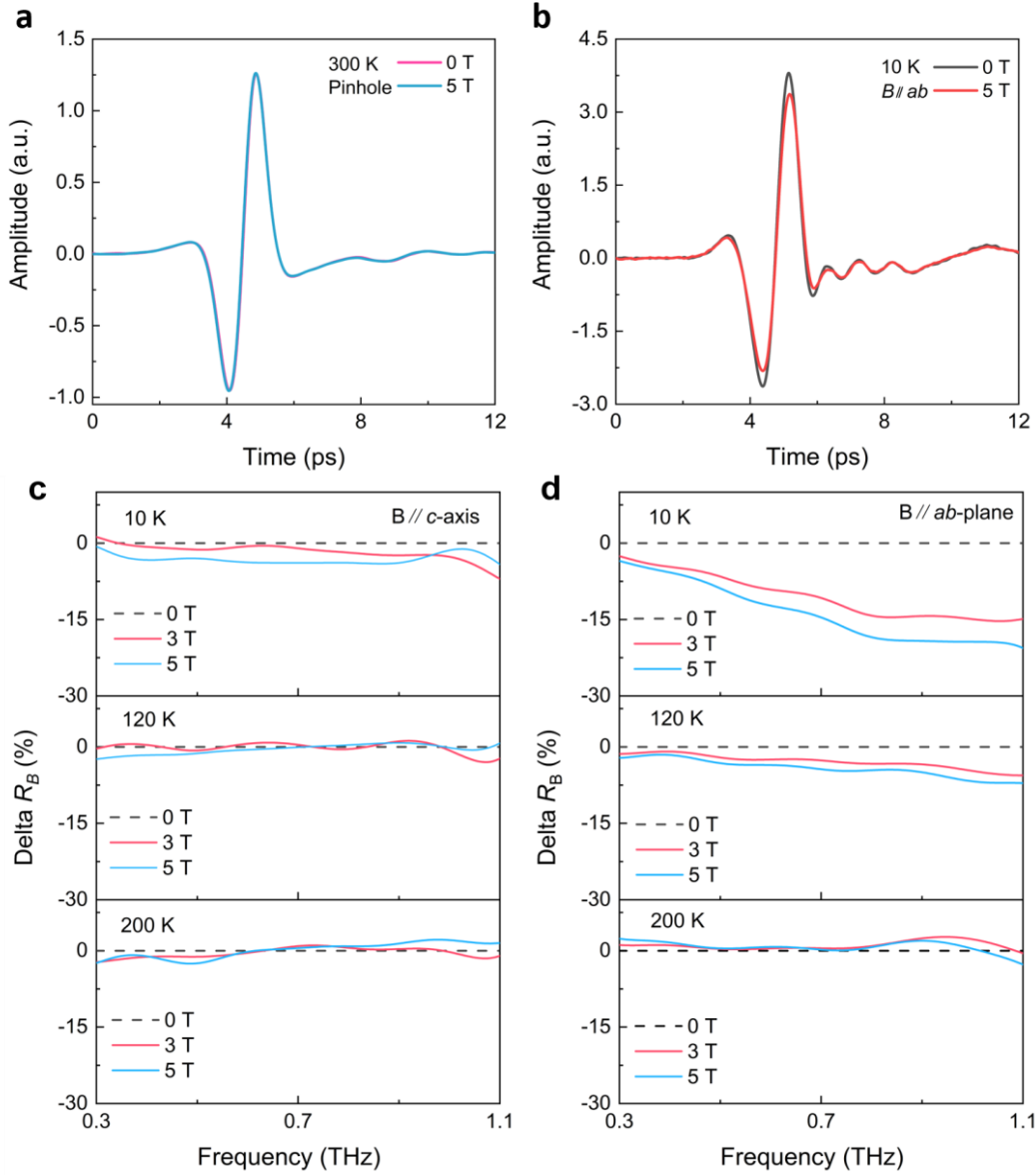

**Fig. S7 | System stability verification and terahertz response of Cr$_2$Ge$_2$Te$_6$ under external magnetic field (*B*). a**, The red and blue curves are the time-resolved waveforms penetrating the pinhole of sample holder without and with magnetic field of 5 T applied. The verification was performed at 300 K. **b**, At 10 K, Time resolved THz responses of Cr$_2$Ge$_2$Te$_6$ without and with *B* of 5 T applied along the *ab*-plane. **c**, At 10 K, 120 K, and 200 K, frequency domain spectral of *Delta R*$_B$ with 0 T, 3 T, and 5 T

magnetic fields applied along the $c$-axis. **d**, Frequency domain spectral of *Delta R*$_B$ with magnetic field along *ab*-plane and temperature same as **c**. The ***B*** comes to affect the amplitude of the radiation, especially in the case along *ab*-plane. While no obvious shift of $f_c$ was observed.